\documentclass{IEEEtran}

\usepackage{graphicx}
\usepackage{caption}
\usepackage{multirow}
\usepackage{textcomp}
\usepackage{marginnote}
\usepackage{booktabs}

\usepackage[margin=1in]{geometry}

\makeatletter %
\let\proof\@undefined
\let\endproof\@undefined
\makeatother %

\usepackage{color}
\usepackage{graphicx}
\usepackage{amssymb}
\usepackage{amsthm}
\usepackage{amsmath}
\usepackage{amsfonts}
\usepackage[ruled]{algorithm}
\usepackage[noend]{algorithmic}
\usepackage{balance}
\usepackage{textcomp}
\usepackage{url}
\usepackage{enumitem} 
\usepackage{comment}
\usepackage[labelfont=bf]{caption} 

\newlength{\figwidths}
\newlength{\figwidthd}
\newlength{\expwidths}
\newlength{\expwidthd}
\theoremstyle{plain}

\makeatletter  
\def\@endtheorem{\hfill\ensuremath{\blacksquare}\endtrivlist\@endpefalse } 
\makeatother
\theoremstyle{definition}

\theoremstyle{remark}


\newcommand{\nop}[1]{}

\begin{document}

\title{Distance-based Data Cleaning: A Survey (Technical Report)}


\author{
Yu Sun$^{1}$,
Jian Zhang$^{2}$\\
$^{1}${School of Software, Tsinghua University, Beijing, China}\\
$^{2}${Shenzhen Research Institute, Peking University, Shenzhen, China}
}

\maketitle

\pagestyle{empty}

\begin{abstract}
\textnormal{With the rapid development of the internet technology,
dirty data are commonly observed in various real scenarios,
e.g., owing to unreliable sensor reading, transmission and collection from heterogeneous sources.
To deal with their negative effects on downstream applications,
data cleaning approaches are designed to preprocess the dirty data before conducting applications.
The idea of most data cleaning methods is to identify or correct dirty data, 
referring to the values of their neighbors which share the same information.
Unfortunately,
owing to data sparsity and heterogeneity, 
the number of neighbors based on equality relationship is rather limited, 
especially in the presence of data values with variances.
To tackle this problem,
distance-based data cleaning approaches propose to consider similarity neighbors based on value distance.
By tolerance of small variants, the enriched similarity neighbors can be identified and used for data cleaning tasks.
At the same time, distance relationship between tuples is also helpful to guide the data cleaning, which contains more information and includes the equality relationship.
Therefore, distance-based technology plays an important role in the data cleaning area, 
and we also have reason to believe that distance-based data cleaning technology will attract more attention in data preprocessing research in the future.
Hence this survey provides a classification of four main data cleaning tasks,
i.e., rule profiling, error detection, data repair and data imputation, 
and comprehensively reviews the state of the art for each class.
}
\end{abstract}



\section{Introduction}
\label{sect:introduction}

Dirty data are prevalent in real applications,
e.g., query answering, data integration, classification and mining.
To deal with the negative effects of dirty data,
data cleaning approaches are considered before conducting data applications.
Data cleaning is the process of detecting and correcting (or removing) corrupt or inaccurate data, 
and refers to identifying incomplete, incorrect, inaccurate or irrelevant parts of the data and then replacing, modifying or deleting the dirty data \cite{DBLP:journals/ress/Wu13}.
The general idea of most existing data cleaning methods is to identify or correct dirty data, 
referring to their neighbors which share the same information.
Unfortunately,
since the data obtained from merging heterogeneous sources often have various representation conventions with variances,
existing equality based data cleaning approaches are not robust enough to address the sparsity and heterogeneity issues.

In order to get more available neighbors providing useful information and be more tolerant to the heterogeneous data,
distance-based data cleaning approaches propose to consider similarity neighbors based on value distance.
By tolerance of small variants, the enriched similarity neighbors can be identified even over the sparse and heterogeneous data,
Moreover, distance relationship between tuple values is also valuable to be an additional signal for data cleaning approaches, 
which indeed includes the equality relationship.
This valuable information can guide the further use-cases.
Therefore, in this survey,
we focus on four primary and important data cleaning tasks,
i.e., rule profiling, error detection, data repair and data imputation, 
and comprehensively reviews the state of the art for each class. 

\begin{figure}[t]
\centering
\includegraphics[width=\linewidth]{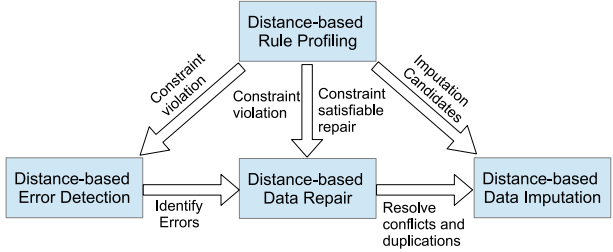}
\caption{
Overview of four distance-based data cleaning tasks
}
\label{fig-overview}
\end{figure}

Figure \ref{fig-overview} provides an overview of these four distance-based data cleaning tasks, with the relationship among them.
As shown, rule profiling is beneficial to all the other three tasks.
It can not only help both distance-based error detection and data repair approaches detect constraint violations,
but also generates available imputation candidates w.r.t. the similar neighbors.
More importantly, any invalid repair/imputation would be abandoned which violates the requirement of rules.
At the same time, distance-based error detection approaches can also help identify possible erroneous cells for the data repair task.
Then the distance-based data repair methods can directly repair these cells,
to avoid excessive repair and information loss.
Moreover, as discussed in \cite{DBLP:series/synthesis/2012Fan}, since the data repair methods can help resolving conflicts and duplications, 
the imputation for missing values can be enriched and improved. 
Therefore, the distance-based data imputation approaches could be performed after the data repair task.

\subsection{Use Cases for Data Cleaning}

Data cleaning has many applicable use-cases, 
including the data management, machine learning and big data analytics scenarios.
We describe several frequently-used use-cases, such
as query answering, data integration, data classification and data mining.

\subsubsection*{Query Answering} 

%

A common assumption in data management is that databases can be kept consistent, that is, satisfying certain desirable integrity constraints.
Unfortunately, owing to various reasons, a dirty database could be inconsistent with respect to the given integrity constraints \cite{DBLP:conf/sigmod/LianCS10}.
Actually, most likely one will be forced to keep using the inconsistent database, because there is still useful information in it.
Then the challenge consists in retrieving the only consistent information from the dirty database.
In consequence, an alternative approach to data cleaning consists in basically living with the inconsistent data,
but making sure that the consistent data can be identified if desired.
When the dirty database answers user queries,
in order to guarantee the correctness of query results,
it can return the only answers that are semantically correct, i.e., that are consistent with the requirement of integrity constraints.
This is the problem of consistent query answering (CQA) \cite{DBLP:series/synthesis/2011Bertossi}.

The consistent data in an inconsistent database is usually characterized as the data that persists across all the database instances that are consistent and minimally differ from the inconsistent instances. 
Those are the so-called repairs of the database. 
In particular, the consistent answers to a query posed to the inconsistent database are those answers that can be simultaneously obtained from all the database repairs.

\subsubsection*{Data Integration} 
Data cleaning and data integration are two important preprocessing steps in data science area.
Data integration is the process of combining data residing in different sources and providing users with a unified view of them \cite{DBLP:conf/pods/Lenzerini02}.
This process becomes significant in a variety of situations, which include both commercial and scientific areas. 
With the rapid development of the computer science,
data integration appears with increasing frequency as the volume and the need to share existing data explodes \cite{lane2006idc}.

At the same time, there is also a huge challenge when integrating data from heterogenous
sources with dirty data. 
The noise and errors in dirty data could prevent the data integration performing, and even mislead the process.
Effective data cleaning over the raw dirty data is definitely helpful to the further data analysis and applications.
After cleaning, the repaired data could be more close to the ground truth, and the data integration over them would be more accurate and reliable.

\subsubsection*{Data Classification}

In the current era when artificial intelligence (AI) technology is sweeping the world,
data classification is the most popular and important topic.
Data classification is the process of organizing data by relevant categories so that it may be used and protected more efficiently,
which involves tagging data to make it easily searchable and trackable. 

It has been widely recognized that the quality of given training data affects the corresponding classification model performance, 
and data scientists spend considerable amount of time on data cleaning before model training \cite{DBLP:journals/corr/abs-1904-09483}.
The quality of classification application depends on the quality of the data it trained on, 
and data cleaning has been the cornerstone of building high-quality classification models.
Accurate data cleaning over the dirty data can improve the quality of data by imputing missing values and repairing erroneous values,
then the classification model performed over high quality data could obtain positive impact,
leading to a better classification accuracy.

\subsubsection*{Data Mining}

In order to get a better known of the given datasets,
data mining technology is often utilized for data analysis.
It is the process of discovering patterns in the datasets and transforming the information into a comprehensible structure for further use, 
which is an interdisciplinary subfield of computer science and statistics \cite{chakrabarti2006data}.

It has been recognized that error data could prevent clustering performing \cite{DBLP:journals/pvldb/GuptaKLMV17,DBLP:conf/kdd/SongLZ15}.
Real data are often dirty with a large number of inaccurate points.
If we directly apply the existing data clustering approaches,
simply discarding a large number of erroneous or incomplete points could obviously affect clustering results,
with too much information loss.
If we first apply data cleaning approaches to preprocess the dirty data,
the data clustering performance could be improved with higher quality data.
Furthermore, \cite{DBLP:conf/kdd/SongLZ15} shows that dirty data could be cleaned and utilized as strong supports in clustering,
which proposes to consider a problem of clustering and repairing over dirty data at the same time.

\subsection{Paper Organization and Contributions}

Data cleaning is an important and practical topic that is closely connected to several other data management and application areas. 
It is also a challenging topic and is becoming increasingly important given the recent trends in data science and big data analytics. 
While there already exist a large body of works that directly design the data cleaning methods referring to the value equality,
the distance-based methods still attract little attention in data cleaning area.
The goal of this survey is to classify and describe four main parts of this work and illustrate the relationship among them.

The remainder of this paper is organized as follows. 
In Sections \ref{sect:rule-profile}, \ref{sect:error-detection}, \ref{sect:data-repair} and \ref{sect:data-imputation}, we survey the state of the art of four main research areas, i.e., rule profiling, error detection, data repair and data imputation, 
of distance-based data cleaning.
Finally, Section \ref{sect:conclusion} concludes this survey.

\section{Distance-based Rule Profiling}
\label{sect:rule-profile}

Data profiling refers to the activity of creating small but informative summaries of a database \cite{DBLP:reference/db/2009},
which is important for analyzing and capturing the data characteristics of real datasets.
Typically, data statistics are translated into rules that are then enforced in a subsequent application phase.
Accurate rule profiling of data can definitely help the further use-cases,
e.g., data cleaning and data integration.

Functional dependencies (FDs) are first used to profile the data semantics.
However, owing to the limited equality relationship, it is difficult to tackle the variety of data (especially the Web data with various information formats).
We thus consider distance-based rules and their corresponding profiling methods in this proposal,
which are listed in Table \ref{table-rule}.

\begin{table}
\begin{tabular}{cccccccccccc}
\toprule
\textsf{Rule} & \textsf{Definition} & \textsf{Discovery} & \textsf{Attributes} \\ \midrule
MDs & \cite{DBLP:conf/pods/Fan08,DBLP:journals/vldb/FanGJLM11} & \cite{DBLP:conf/cikm/SongC09,DBLP:journals/tkdd/WangSCYC17,DBLP:journals/dke/Song013} & LHS \\
MFDs & \cite{DBLP:conf/icde/KoudasSSV09} & \cite{DBLP:conf/icde/KoudasSSV09} & RHS \\
CDs & \cite{DBLP:conf/icde/SongCY11,DBLP:journals/vldb/Song0Y13} & \cite{DBLP:journals/vldb/Song0Y13} & LHS\&RHS \\
FFDs & \cite{DBLP:journals/tods/RajuM88} & \cite{DBLP:conf/iceb/WangC04a,DBLP:books/igi/Galindo08/WangSH08} & LHS\&RHS \\ 
DDs & \cite{DBLP:journals/tods/Song011} & \cite{DBLP:journals/tods/Song011,DBLP:conf/icde/SongCC12,DBLP:journals/tkde/Song0C14,DBLP:conf/adc/KwashieLLY14} & LHS\&RHS \\
\bottomrule
 \end{tabular}
\caption{Representative distance-based rules with different features}
\label{table-rule}
\end{table}

\subsection{Matching Dependencies (MDs)}

Matching dependencies (MDs) are proposed in \cite{DBLP:conf/pods/Fan08,DBLP:journals/vldb/FanGJLM11} to specify matching rules for object identification,
a.k.a, data deduplication, record linkage, merge-purge and record matching.
Given one or more relations, MDs help us to identify tuples from those relations that refer to the same real-world object.
Relative candidate keys (RCKs) \cite{DBLP:journals/pvldb/Song0C14} are used to specify which attributes to compare and how to compare them in order to identify those tuples related to the same object.

Consider a relation instance $\mathit{I}$ with schema $\mathcal{R}=(\mathit{A}_{1},\dots,\mathit{A}_{m})$.
Let $\mathsf{dom}(A_{j})$ denote all the values of an attribute $\mathit{A}_{j}\in\mathcal{R}$ in $\mathit{I}$, i.e., $\mathsf{dom}(A_{j})=\Pi_{A_{j}}(\mathit{I})$.

An MD defined on schema $(\mathcal{R}_{1},\mathcal{R}_{2})$ is an expression of the form
$$
(\mathcal{R}_{1}(\mathit{X}_{1})\approx\mathcal{R}_{2}(\mathit{X}_{2}))
\rightarrow
(\mathcal{R}_{1}(\mathit{Y}_{1})\rightleftharpoons\mathcal{R}_{2}(\mathit{Y}_{2})),
$$
where (1) $(\mathit{X}_{1},\mathit{X}_{2})$ and $(\mathit{Y}_{1},\mathit{Y}_{2})$ are sets of compatible attributes over $(\mathcal{R}_{1},\mathcal{R}_{2})$; (2) $\approx$ is the similar operator indicating that the values are similar; (3) $\rightleftharpoons$ is the matching operator on attributes $\mathit{Y}_{1}$ and $\mathit{Y}_{2}$, denoting that the values are identified.
The MD implies that for any two tuples from instances of $\mathcal{R}_{1}$ and $\mathcal{R}_{2}$,
if they have similar values between $\mathcal{R}_{1}(\mathit{X}_{1})$ and $\mathcal{R}_{2}(\mathit{X}_{2})$,
their values on $\mathcal{R}_{1}(\mathit{Y}_{1})$ and $\mathcal{R}_{2}(\mathit{Y}_{2})$ should be identified.

An exact MDs discovering algorithm with certain utility requirements of support and confidence is proposed in \cite{DBLP:conf/cikm/SongC09,DBLP:journals/tkdd/WangSCYC17},
together with pruning strategies to improve the time performance.
Since the exact algorithm has to traverse all the data during the discovering, 
\cite{DBLP:journals/dke/Song013} considers an approximate method which only uses a part of the data,
with an approximation bound of relative errors on support and confidence of returned MDs.
The approximation algorithm only traverses the first $k$ tuples in statistical distribution, instead of the whole dataset. 
Moreover, RCKs with minimal compared attributes can remove redundant semantics.

\subsection{Metric Functional Dependencies (MFDs)}

Metric functional dependencies (MFDs) \cite{DBLP:conf/icde/KoudasSSV09} propose to extend the traditional functional dependencies with distance metrics on RHS attributes,
in order to tackle the small variations in data format and interpretation.
MFDs allow small differences of the heterogeneous data,
which are obtained from merging various sources.

A metric functional dependency (MFD) is defined in the form of
$$
X\stackrel{\delta}{\longrightarrow} A,
$$
where (1) $\mathit{X}$ is a set of attributes and $A$ is a single attribute in $\mathcal{R}$,
(2) $\delta$ is a distance tolerance parameter of the distance metric $d$ on attribute $A$.
The distance metric $d$ is defined on the domain of $A$, having $\textsf{dom}(A)\times\textsf{dom}(A)\rightarrow\mathit{R}$.
Given a relation instance $\mathit{I}$ with schema $\mathcal{R}$, 
left-hand-side (LHS) attributes $X\subset\mathcal{R}$ and right-hand-side (RHS) attribute $A\in\mathcal{R}$, a distance metric $d$ defined over $A$, and a parameter $\delta\geq0$,
the MFD $X\stackrel{\delta}{\longrightarrow} A$ is said to hold if any two tuples $\mathit{t}_{1},\mathit{t}_{2}\in\mathit{I}$ having $\mathit{t}_{1}[X]=\mathit{t}_{2}[X]$ must have distance $\leq\delta$ on attribute $A$.

Given a relational instance $\mathit{I}$, 
in order to discover MFDs from it,
an important step is to verify whether a candidate MFD holds over instance $\mathit{I}$.
In the discovery procedure, 
it first groups all the tuples according to LHS attributes $X$, and the maximum distance between each pair of tuples is computed, recorded as the diameter.
The problem of finding appropriate parameters for which the dependency holds can be reduced to the verification problem via binary search.
For general metrics,
the verification procedure costs $O(n^2)$ to assure whether the candidate MFD holds,
and it costs $O(n)$ to verify whether there exists $\delta'\leq2\delta$ such that $X\stackrel{\delta}{\longrightarrow} A$ holds.
Moreover, for euclidean metrics,
the diameter computation is trivial,
which can be performed in $O(n\log\mathit{n})$ time by computing the convex hull of the points and walking along the boundary using rotating calipers method.
For string metrics, the cosine similarity of $q$-gram vectors is usually used as the distance metric.
The given MFD holds if the diameter exists in the given relation instance.

\subsection{Comparable Dependencies (CDs)}


Comparable dependencies (CDs) \cite{DBLP:conf/icde/SongCY11,DBLP:journals/vldb/Song0Y13} specify constraints on comparable attributes,
in order to capture the data dependencies of two relations from heterogenous sources.

A comparable function between two attributes $\mathit{A}_{i}$ and $\mathit{A}_{j}$ is defined as follows
$$
\theta(\mathit{A}_{i},\mathit{A}_{j}):[\mathit{A}_{i}\leftrightarrow_{ii}\mathit{A}_{i},\mathit{A}_{i}\leftrightarrow_{ij}\mathit{A}_{j},\mathit{A}_{j}\leftrightarrow_{jj}\mathit{A}_{j}],
$$
which specifies a constraint on the comparable relationship of two values from attribute $\mathit{A}_{i}$ or $\mathit{A}_{j}$,
according to their corresponding comparison operators $\leftrightarrow_{ii}$, $\leftrightarrow_{ij}$ or $\leftrightarrow_{jj}$.
The comparison operator $\leftrightarrow_{ij}$ could be an equality metric or matching operator.
For any two tuples $\mathit{t}_{1}$ and $\mathit{t}_{2}$ in the given dataset, we say that $\mathit{t}_{1}$ and $\mathit{t}_{2}$ agree on a comparison function, denoted by $(\mathit{t}_{1},\mathit{t}_{2})\asymp\theta(\mathit{A}_{i},\mathit{A}_{j})$, if at least one pair of $(\mathit{t}_{1}[\mathit{A}_{i}],\mathit{t}_{2}[\mathit{A}_{i}])$, 
$(\mathit{t}_{1}[\mathit{A}_{i}],\mathit{t}_{2}[\mathit{A}_{j}])$,
$(\mathit{t}_{1}[\mathit{A}_{j}],\mathit{t}_{2}[\mathit{A}_{i}])$,
$(\mathit{t}_{1}[\mathit{A}_{j}],\mathit{t}_{2}[\mathit{A}_{j}])$ agrees on the corresponding comparison operator specified by $\theta(\mathit{A}_{i},\mathit{A}_{j})$.
That is, the two tuples $\mathit{t}_{1}$ and $\mathit{t}_{2}$ are said to be similar w.r.t. $\theta(\mathit{A}_{i},\mathit{A}_{j})$, 
if at least one the of three similarity operators evaluates to true.
Otherwise, it is denoted by $(\mathit{t}_{1},\mathit{t}_{2})\not\asymp\theta(\mathit{A}_{i},\mathit{A}_{j})$

A comparable dependency (CD) with comparison functions has the form of
$$\mathsf{CD}:\bigwedge\theta(\mathit{A}_{i},\mathit{A}_{j})\rightarrow\theta(\mathit{B}_{p},\mathit{B}_{q}),$$
where $\theta(\mathit{A}_{i},\mathit{A}_{j})$ and $\theta(\mathit{B}_{p},\mathit{B}_{q})$ are comparison functions. 
The LHS and RHS attributes of CD are denoted by $\bigwedge\theta(\mathit{A}_{i},\mathit{A}_{j})$ and $\theta(\mathit{B}_{p},\mathit{B}_{q})$ respectively.
It states that for any two tuples $\mathit{t}_{1}$ and $\mathit{t}_{2}$ that are comparable w.r.t. $\theta(\mathit{A}_{i},\mathit{A}_{j})$,
it implies $(\mathit{t}_{1},\mathit{t}_{2})\asymp\theta(\mathit{B}_{p},\mathit{B}_{q})$.

In order to efficiently discover CDs from the given datasets,
\cite{DBLP:journals/vldb/Song0Y13} proposes a pay-as-you-go profiling approach.
The algorithm conducts an incremental discovery of comparable dependencies w.r.t. the new identified comparison functions.

\subsection{Fuzzy Functional Dependencies (FFDs)}
Since FDs can only take care of well-defined and unambiguous data with the strict equality relationship,
Fuzzy functional dependencies (FFDs) \cite{DBLP:journals/tods/RajuM88} are proposed to be more applicable with real data,
which are often partially known incomplete or imprecise.
Under such cases,
there is no clear way to check whether two incomplete/imprecise values are equal,
and the traditional functional dependency cannot be directly applied.
FFDs extend the classical FDs to deal with fuzzy information,
which consider integrity constraints that may involve fuzzy concepts.
In the fuzzy domain, the equality of domain values defines a fuzzy proposition and is specified as ``approximately equal'', ``more or less equal'', etc.

For each attribute $\mathit{A}_{j}\in\mathcal{R}$,
a fuzzy resemblance relation EQUAL(EQ) is defined to be a subset of $\mathsf{dom}(A_{j})\times\mathsf{dom}(A_{j})$,
where $\mu_{EQ}$ satisfies the reflexivity and symmetry conditions.
In terms of possibility theory,
$\mu_{EQ}(\mathit{a},\mathit{b}),\mathit{a},\mathit{b}\in\mathsf{dom}(A_{j})$ can be interpreted as the possibility of treating $\mathit{a}$ and $\mathit{b}$ as approximately equal.
That is,
the larger the $\mu_{EQ}(\mathit{a},\mathit{b})$ is,
the more possibility the values $\mathit{a}$ and $\mathit{b}$ are ``equal''.
Then the definition of EQUAL can be extended over all attributes $\mathcal{R}$,
with the membership function
\begin{align*}
\mu_{EQ}(\mathit{t}_{1},\mathit{t}_{2})=
\min\{&\mu^{1}_{EQ}(\mathit{t}_{1}[\mathit{A}_{1}],\mathit{t}_{2}[\mathit{A}_{1}]),
\\
&\mu^{2}_{EQ}(\mathit{t}_{1}[\mathit{A}_{2}],\mathit{t}_{2}[\mathit{A}_{2}]),\dots,
\\
&\mu^{m}_{EQ}(\mathit{t}_{1}[\mathit{A}_{m}],\mathit{t}_{2}[\mathit{A}_{m}])\}
\end{align*}

Then a fuzzy functional dependency (FFD) $\mathit{X}\rightsquigarrow\mathit{Y}$ with $\mathit{X},\mathit{Y}\subset\mathcal{R}$, holds in a fuzzy relation instance $\mathit{I}$,
if for all tuples $\mathit{t}_{1}$ and $\mathit{t}_{2}$ of $\mathit{I}$, we have
\begin{align*}
\mu_{EQ}(\mathit{t}_{1}[\mathit{X}],\mathit{t}_{2}[\mathit{X}])\leq
\mu_{EQ}(\mathit{t}_{1}[\mathit{Y}],\mathit{t}_{2}[\mathit{Y}]).
\end{align*}
$\mu_{EQ}(\mathit{t}_{1}[\mathit{X}],\mathit{t}_{2}[\mathit{X}])$ denotes the fuzzy resemblance relation of tuples $\mathit{t}_{1}$ and $\mathit{t}_{2}$ on attributes $\mathit{X}\subset\mathcal{R}$, and $\leq$ means that the possibility of the equality between tuples $\mathit{t}_{1}$ and $\mathit{t}_{2}$ on attributes $\mathit{X}$ is less than that on $\mathit{Y}$.
It means that the values of tuples $\mathit{t}_{1}$ and $\mathit{t}_{2}$
on attributes $\mathit{Y}$ have more possibility to be “equal” than those on attributes $\mathit{X}$.

It is notable that an FD can be viewed as a special case of an FFD: $\mathit{X}\rightsquigarrow\mathit{Y}$.
To prove this, we can suppose that $\mu_{EQ}(\mathit{a},\mathit{b}),\mathit{a},\mathit{b}\in\mathsf{dom}(A_{j})$, satisfies the additional property that $\mu_{EQ}(\mathit{a},\mathit{b})=0$ for $\mathit{a}\neq\mathit{b}$.
Then the dependency $\mathit{X}\rightsquigarrow\mathit{Y}$ implies that no two tuples of $\mathit{I}$ can agree in $\mathit{X}$ attribute values,
yet disagree in their $\mathit{Y}$ attribute values.

Following \cite{DBLP:conf/iceb/WangC04a},
users can automatically mine the FFDs from given dataset,
which is an extension of the TANE algorithm \cite{DBLP:conf/icde/HuhtalaKPT98,DBLP:journals/cj/HuhtalaKPT99} for mining FDs.
\cite{DBLP:conf/iceb/WangC04a} is deliberated to find non-trivial FFDs,
each with a single attribute in its right-hand-side. 
The mining algorithm checks
every two tuples to see if it satisfies the EQUAL relation.
Moreover, 
in order to optimize the rule discovering efficiency,
\cite{DBLP:books/igi/Galindo08/WangSH08} considers an incremental searching solution based on pair-wise comparison,
which avoids re-scanning of the whole dataset for the new added tuples.


\subsection{Differential Dependencies (DDs)}

Differential dependencies (DDs) \cite{DBLP:journals/tods/Song011} propose a more advanced differential function to express
the constraints on distances, which can further capture the distance relationship between tuples. 
If two tuples have distances on LHS attributes $\mathit{X}$ agreeing with a certain differential function, 
then their distances on RHS attribute $\mathit{A}$ should also agree with another differential function. 
DDs can express constraints not only on similarity but also on dissimilarity.

For each attribute $A\in\mathcal{R}$, DDs associate it with a distance metric $\mathit{d}_A$,
%
which can be any distance evaluation function such as \cite{DBLP:journals/isci/SongZ014}.
For instance,
it can be the absolute value of difference on a numerical attribute or edit distance for a categorical attribute.

A differential function $\phi[A]$ on attribute $A$ specifies a distance restriction by a range of metric distances over $A$. 
We say that two tuples $t_1,t_2$ in a relation $\mathit{I}$ are compatible w.r.t.\ the differential function $\phi[A]$, denoted by $(t_1,t_2)\asymp\phi[A]$ or $(t_1[A],t_2[A])\asymp\phi[A]$, if the metric distance of $t_1$ and $t_2$ on attribute $A$ is within the range specified by $\phi[A]$, a.k.a. satisfy with the distance restriction $\phi[A]$.
As the metric is symmetric, it is equivalent to write  $(t_2,t_1)\asymp\phi[A]$.

A differential function may also be specified on a set of attributes $X$, say $\phi[X]$, which denotes a pattern of differential functions (distance ranges) on all the attributes in $X$. 
$\phi[A]$ is a projection on attribute $A$ of $\phi[X], A\in X$.

A DD over $\mathcal{R}$ has a form $(X\rightarrow A, \phi[XA])$
where $X\subseteq\mathcal{R}$ are determinant attributes, $A\in\mathcal{R}$ is the dependent attribute, and $\phi[XA]$ is a differential function on attributes $X$ and $A$.
It states that any two tuples from $\mathcal{R}$ satisfying the differential function $\phi[X]$ must satisfy $\phi[A]$ as well, $\phi[X]$ and $\phi[A]$ are the projections of differential function $\phi[XA]$ on $X$ and $A$, respectively.

A relation $\mathit{I}$ of $\mathcal{R}$ \emph{satisfies} a DD, denoted by $\mathit{I}\vDash (X\rightarrow A,\phi[XA])$, if for any two tuples $t_1$ and $t_2$ in $\mathit{I}$, $(t_1,t_2)\asymp\phi[X]$ implies $(t_1,t_2)\asymp\phi[A]$.
We say a relation $\mathit{I}$ satisfies a set $\Sigma$ of DDs, $\mathit{I}\vDash\Sigma$, if $\mathit{I}$ satisfies each DD in $\Sigma$. 

In order to design the DDs profiling algorithm,
\cite{DBLP:journals/tods/Song011} introduces the concept of minimal DDs,
which can be discovered by finding a minimal cover set of DDs that holds in a given relation.
Since discovering a minimal cover is proven to be NP-hard,
several pruning strategies \cite{DBLP:conf/icde/SongCC12,DBLP:journals/tkde/Song0C14} are then devised for accelerating DDs profiling,
where the determination of distance thresholds for differential functions is studied in a parameter-free style. 
Moreover,
in order to avoid the mining of an extremely large set,
\cite{DBLP:conf/adc/KwashieLLY14} proposes a subspace clustering-based approach to discover DDs.

\section{Distance-based Error Detection}
\label{sect:error-detection}


%
Distance-based error detection methods often identify the tuples containing values significantly distant from the others \cite{hawkins1980identification},
which are referred as abnormalities, discordants, deviants and anomalies.
As such, outlying points usually have greater suspicion to be generated by different mechanisms from the other normal points, 
and are more suspicious to be errors.

Table \ref{table-detection} summarizes the representative distance-based error detection methods,
which are mainly categorized into two categories,
i.e., based on distance clustering and distance distribution.

\begin{table}[t]
 \centering
 \begin{tabular}{cccccc}
\toprule  
Method & Category & Signal Type\\ 
\midrule 
DBSCAN \cite{DBLP:conf/kdd/EsterKSX96} &  \multirow{3}{1.2cm}{clustering} & local  \\
OPTICS \cite{DBLP:conf/sigmod/AnkerstBKS99} & & local  \\
LS \cite{DBLP:journals/pvldb/GuptaKLMV17}  &  & local  \\
\hline
DBO \cite{knorr1998algorithms} & \multirow{4}{1.2cm}{distribution} & global  \\
dBoost \cite{mariet2016outlier} & & global  \\
LOF \cite{DBLP:conf/sigmod/BreunigKNS00} & & local \\ 
Subspace \cite{DBLP:conf/icdm/MicenkovaNDA13} &  & local \\ 
 \bottomrule
\end{tabular}
 \caption{Representative distance-based error detection methods with different features}
\label{table-detection}
\end{table}

\subsection{Distance Clustering based}

In density-based clustering, clusters are defined as areas with higher density \cite{DBLP:reference/db/Ester18,DBLP:journals/widm/KriegelKSZ11}.
For instance, in DBSCAN \cite{DBLP:conf/kdd/EsterKSX96},
a cluster
is a set of density-connected points which
is maximal with respect to density-reachability.
And OPTICS \cite{DBLP:conf/sigmod/AnkerstBKS99}
orders the clusters with
respect to higher density with lower distance,
when the intrinsic cluster structure cannot be characterized by a global density,

Specifically, DBSCAN \cite{DBLP:conf/kdd/EsterKSX96} considers a clustering where each cluster is a set of density-connected points which is maximal with maximal density. 
It proposes to classify all the tuples into three kinds based on their distances between each other.
The tuples with sufficient near neighbors will be core points, and the neighbors of them without enough neighbors are regraded as border points.
The density-based clustering is performing over the core points and border points,
which are density-reachable with each other.
Those tuples which are neither core points nor border points will be detected as errors.
That is, the core points and border points are clustered, 
and all the other points will be identified as erroneous.

Since the intrinsic cluster structure of many datasets could not be characterized by global density in some real scenarios,
OPTICS \cite{DBLP:conf/sigmod/AnkerstBKS99} proposes to utilize the local density to capture the clusters in different regions of the given dataset.
Actually, OPTICS seems like an extended DBSCAN algorithm for an infinite number of distance threshold.
The difference is that OPTICS does not assign cluster memberships for each point.
Instead, it creates an augmented ordering of the database representing its density-based clustering structure,
and stores the information of the objects processing order and which would be used.
For each point, the information consists of the core-distance and reachability-distance,
which are used to extract all density-based clusters.

LS \cite{DBLP:journals/pvldb/GuptaKLMV17} performs $k$-means clustering with errors,
which aims to cluster all the points to minimize the variance of those assigned to the same cluster with ignoring a set of possible error points.
LS proposes a local search-based algorithm based on $k$-means clustering to find the abnormal points,
with performance guarantee w.r.t. the clustering quality.
It iteratively checks if swapping one of the current clustering centers with a non-center and any further local swap with the additional removal of errors could improve the clustering performance.
If it is the case, the swap and removal will be performed.
This procedure continues until there is no possible operation that could lead to a significant improvement of the clustering performance.

\subsection{Distance Distribution based}

Distance based detection approaches
\cite{knorr1998algorithms,DBLP:journals/vldb/KnorrNT00} determine a fraction $\mathit{p}$ and distance threshold $\epsilon$ according to data distributions.
Then a tuple $\mathit{t}_{i}$ in a relation instance $\mathit{I}$ will be identified as an error if at least fraction $\mathit{p}$ of the tuples in $\mathit{I}$ lies greater than a distance threshold $\mathit{\epsilon}$ from $\mathit{t}_{i}$.
Given $\mathit{p}$ and $\mathit{\epsilon}$,
the simplest method to find all the errors in the given dataset is to answer a nearest neighbor or range query centered at each tuple $\mathit{t}_{i}$.
In order to efficiently find all the errors,
DBO \cite{knorr1998algorithms} presents a cell-based algorithm with a linear complexity of the number of tuples in $\mathit{I}$ but exponential with the number of attributes.
The experimental results in \cite{DBLP:journals/vldb/KnorrNT00} show that the proposed solution achieves a good performance in real-life applications.

dBoost \cite{mariet2016outlier} relies on several classical error detection approaches,
such as histogram, gaussian modeling and multivariate gaussian mixture models,
to analyze the data distributions for error detection.
It applies tuple expansion features to expand string values into numerical values and find string errors as well.
In this sense, dBoost can detect errors from both the numerical data and the categorical data.
Moreover, dBoost reconstructs structured and semantically rich information from raw data using a library of expansion rules,
which can be fed to both single and multi-variate models to detect errors.

LOF \cite{DBLP:conf/sigmod/BreunigKNS00} uses the $k$-distance of each tuple to measure its local density-based error factor.
The error factor of a tuple captures the degree to which it is called an error.
This factor is in local degree which depends on how isolated the object is with respect to the surrounding neighborhood.
The lower a tuple's $k$-distance is, and the higher is the local error factor.
Any tuple with significantly lower density will be identified as errors.
Moreover, LOF also has some properties including the lower and upper bound for any object,
which are even tight for some classes of objects.


Tuples with erroneous values usually tend to be outlying with others, 
which are extraordinary objects in a data collection.
However, given an outlier tuple with possible errors,
a key problem is to know which attributes in this tuple indeed contain errors.
Given a tuple with possible erroneous values which could be identified by the other existing detection algorithms,
Subspace \cite{DBLP:conf/icdm/MicenkovaNDA13} aims to find which attribute values indeed make these tuples deviate from the rest of tuples.
In this sense, it is complementary to the other existing detection algorithms.
\cite{DBLP:conf/icdm/MicenkovaNDA13} designs a scoring function to measure the error degree of subspace attributes,
which is computed by considering the value distances between tuples in the same subspace attributes.
An erroneous subspace for a given tuple is a set of attributes whose score is relatively high,
which will be identified as possible erroneous attributes.

\section{Distance-based Data Repair}
\label{sect:data-repair}
%


\begin{table}[t]
 \centering
 \begin{tabular}{cccccc}
\toprule  
Method & Category & Minimum Change \\ 
\midrule 
DORC \cite{DBLP:conf/kdd/SongLZ15} & clustering  & $\surd$  \\
\hline
srFn \cite{icde20-swapping} & cost & $\times$  \\
\hline
ERACER \cite{DBLP:conf/sigmod/MayfieldNP10} & model & $\times$  \\
\hline
Hybrid \cite{DBLP:journals/pvldb/SongCY014} & \multirow{5}{1.2cm}{constraint} & $\surd$  \\
Grepair \cite{DBLP:journals/vldb/SongLCYC17} & & $\surd$  \\
CVtolerant \cite{DBLP:conf/sigmod/SongZW16} & & $\surd$   \\ 
UniClean \cite{DBLP:conf/sigmod/FanLMTY11} & & $\surd$   \\ 
SDs \cite{DBLP:journals/pvldb/GolabKKSS09} & & $\surd$   \\ 
 \bottomrule
\end{tabular}
 \caption{Representative distance-based data repair methods with different features}
 \label{table-repair}
\end{table}

As far as we know,
few distance-based studies have been proposed to address data repairing problem. 
As shown in Table \ref{table-repair},
we could briefly summarize the typical distance-based data repairing methods into four categories,
and most of them repair the dirty data based on the distance-based constraints.

\subsection{Distance Clustering based}
DORC \cite{DBLP:conf/kdd/SongLZ15} provides a distance clustering based data repairing solution.
The approach simultaneously repairs dirty data w.r.t. the local density of data during
the clustering process,
which treats low density data objects as errors and move (repair) them to high density ones as repairs,
i.e., making each outlier an inlier.
It generally extends the setting of DBSCAN \cite{DBLP:conf/kdd/EsterKSX96}. 
That is, while core points and border points are clustered, 
all the other outlying points will be repaired.
Following the minimum change principle in data repairing \cite{DBLP:journals/tods/Wijsen05}, 
the objective of DORC is to find a minimum repair of data such that all the repaired data can be clustered.
It provides a linear time approximate approach LDORC via grouping data points with performance guarantee,
which achieves a good trade-off between efficiency and effectiveness.

\subsection{Distance Cost based}

srFn \cite{icde20-swapping} designs a distance cost based swapping repair method for repairing misplaced attributes values.
In order to handle the misplaced attribute values,
which are commonly observed and could be introduced generally in all ETL steps,
srFn proposes to evaluate the likelihood of a swapping repaired tuple by studying its distances to neighbors.
Instead of directly evaluating how likely a tuple contains attribute values appearing exactly in the value distribution,
srFn checks whether the tuple has values similar to other tuples,
in order to be tolerant to data sparsity and heterogeneity.
If the tuple is distant from others,
misplaced-attribute errors are likely to occur.
It shows that if considering all the tuples in dataset as neighbors in evaluating a repair, 
the swapping repair problem can be reduced to the minimum weight perfect matching  problem \cite{DBLP:books/daglib/p/Kuhn10} and can be polynomial time solvable by using Hungarian method.
The special case further motivates its approximation algorithm by considering a fixed set of $\mathit{\kappa}$ neighbors,
which are generated by nearest neighbors of each tuple in the dataset.

\subsection{Distance Model based}


ERACER \cite{DBLP:conf/sigmod/MayfieldNP10} considers a convolution model on distances to represent the dependencies in the given dataset with possible erroneous values.
In addition to the specific values between random variables,
the dependencies could be modeled based on the distance of values.
For a pair of attributes that could be correlated between each other,
ERACER proposes to construct the distance distribution between two variables as the dependency.
Comparing to the traditional value distribution model,
ERACER allows to use all the information across the entire dataset for data repairing,
instead of matching on exact combinations of specific attributes.
In this sense, the convolution model on distances is applicable to repair the heterogeneous data.
After the convolution models learning,
ERACER considers each determinant attribute's inference independently and combine the  results of them as the final repair.

\subsection{Distance Constraint based}

Hybrid \cite{DBLP:journals/pvldb/SongCY014} utilizes the neighborhood constraints based distances,
which specify label pairs that are allowed to appear on adjacent vertexes in the graph, 
to repair erroneous vertex labels.
It proposes several approximate repairing algorithms,
including greedy heuristics, contraction method and a hybrid approach,
to make the repaired graphs satisfying neighborhood constraints.
The hybrid approach takes the advantages of both greedy and contraction methods.
The greedy method is utilized first to efficiently eliminate violations,
and contraction process is conducted when no violation could be further reduced. 
These two operations are applied alternatively, where the greedy method has a higher priority.

Since errors may also occur on vertex neighbors and introduce violations to the constraint graph,
Grepair \cite{DBLP:journals/vldb/SongLCYC17} further proposes the extended algorithm for the graph repairing with both vertex label and neighbor repairs.
A cubic-time approximation algorithm with constant-factor performance guarantee is designed to solve this problem,
where one vertex with either label or neighbor is repaired in each operation.
The algorithm conducts iteratively,
and it requires that all the violations to the vertex are eliminated after the repair operation to ensure the termination.

In addition to the erroneous data,
CVtolerant \cite{DBLP:conf/sigmod/SongZW16} notices that the constraints may also be imprecise.
The imprecise constraints could be oversimplified so that correct data are erroneously identified as violations, 
or overrefined that the constraints overfit the data and fail to identify true violations.
Therefore, it proposes a $\theta$-tolerant repair,
which allows a small variation of the constraints when repairing dirty data.
The proposed violation-free repair method puts both violations and suspect data into the repair context,
which ensures no new violation introduced after repairing.
 
UniClean \cite{DBLP:conf/sigmod/FanLMTY11} proposes a uniform framework for data cleaning via both matching and repairing with the help of matching dependencies,
consisting of a three-phase solution with three algorithms.
It designs an algorithm to identify deterministic fixes which are accurate, referring to the confidence analysis and master data.
Another algorithm utilizes the information entropy and inferring evidence to compute reliable fixes, in order to improve the repair performance.
For the other remaining errors, UniClean considers a heuristic-based method to find a repair for fixing the dirty data.
These three methods are complementary to each other, in order to get a higher data repairing accuracy.

In order to better express the semantics of sequential data with ordered domains and help handling data quality problems in such data,
SDs \cite{DBLP:journals/pvldb/GolabKKSS09} proposes the concept of sequential dependencies.
It further considers the dependencies approximately and conditionally,
to make the rules more applicable to the real-world data.
Moreover,
SDs provides a framework to discover pattern tableaux, 
which are compact representations of the subsets data that satisfy the underlying dependencies.

\section{Distance-based Data Imputation}
\label{sect:data-imputation}

The idea of imputing missing values based on other available information, 
e.g., complete attributes or complete tuples,
is widely adopted in solving data imputation problems,
and has been successfully applied in many important areas.
Table \ref{table-imputation} lists the representative distance-based data imputation studies,
which are mainly categorized into four classes with different features.

\begin{table}[t]
 \centering
 \begin{tabular}{llll}
\toprule  
Method & Category & Data Type \\ 
\midrule 
kNN \cite{altman1992introduction} & \multirow{3}{1.2cm}{neighbor}   & string\&number  \\
kNNE \cite{DBLP:conf/icpr/DomeniconiY04} & & string\&number \\
MIBOS \cite{DBLP:conf/smc/WuFHW12} & & string  \\
\hline
CMI \cite{DBLP:journals/tcos/ZhangZZQZ08} & \multirow{3}{1.2cm}{clustering} & string  \\
FKM \cite{li2004towards} &  & string\&number  \\
IFC \cite{DBLP:conf/fuzzIEEE/NikfalazarYBK17} &  & string\&number  \\
\hline
LOESS \cite{Cleveland1996} & \multirow{3}{1.2cm}{model} & number \\
IIM \cite{DBLP:conf/icde/ZhangSSW19} &  & number  \\ 
DLM \cite{kdd20-distanceModel} &  & string\&number  \\ 
\hline
Derand \cite{DBLP:journals/pvldb/SongZC015,DBLP:journals/tkde/SongSZCW20} &  \multirow{2}{1.2cm}{constraint} & string\&number  \\ 
HoloClean \cite{DBLP:journals/pvldb/RekatsinasCIR17} &  & string\&number  \\ 
 \bottomrule
\end{tabular}
 \caption{Representative distance-based data imputation methods with different features}
 \label{table-imputation}
\end{table}

\subsection{Distance Neighbor based}

For each incomplete tuple $\mathit{t}_0$ with missing values $\mathit{t}_0[\mathit{U}]$ on attributes $\mathit{U}$,
distance neighbor based imputation methods consider the tuples with short tuple distances defined over the complete attributes of the incomplete tuple.
kNN \cite{altman1992introduction} finds a set of neighbors $\mathit{t}_i$ for the incomplete tuple $\mathit{t}_0$, 
which have similar values with $\mathit{t}_0$ on its complete attributes $\mathit{t}_0[\mathcal{R}\setminus\mathit{U}]$. 
The values on attribute $\mathit{A}\in\mathit{U}$ of the neighbors are then aggregated as the imputation of the missing value $\mathit{t}_0[\mathit{A}]$.
It can be the majority voting principle for categorical values or simply the mean value for numerical attributes.

Recent ensemble work has shown that combining predictors can lead to significant reduction in generalization error,
compared with utilizing the individual predictors.
However, many combining methods do not improve the kNN method,
since it is robust with respect to variations of a data set. 
In contrast, it is sensitive to the input features.
Therefore,
instead of considering the fixed complete attribute values $\mathcal{R}\setminus\mathit{U}$, 
the kNNE \cite{DBLP:conf/icpr/DomeniconiY04} explores more neighbor distances on various subsets of the complete attributes $\mathcal{R}\setminus\mathit{U}$. 
kNNE exploits the instability of kNN method with respect to different choices of features to generate a diverse neighbor set with possibly uncorrelated errors.
Since the candidates suggested by neighbors may distant with each other, it is also important to consider how to accurately aggregate the values from different similar neighbor sets.
 
MIBOS \cite{DBLP:conf/smc/WuFHW12} considers a tuple distance which is computed by evaluating the attribute values between tuples.
MIBOS proposes a tuple distance model between tuples with incomplete data,
by constructing the distance matrix of tuples and using the nearest neighbors of each incomplete tuple to impute the missing data iteratively.
In the iteration strategy, the filled value in the imputation process will be directly applied to the imputation of the same iteration.

\subsection{Distance Clustering based}

Instead of finding neighbors with nearest distances on various subsets of the complete attributes $\mathcal{R}\setminus\mathit{U}$, 
clustering methods are also employed to explore different groups of neighbors, 
e.g., by kernel function imputation strategy \cite{DBLP:journals/tcos/ZhangZZQZ08},
fuzzy k-means \cite{li2004towards} 
and its iterative manner \cite{DBLP:conf/fuzzIEEE/NikfalazarYBK17}. 

CMI \cite{DBLP:journals/tcos/ZhangZZQZ08} considers clusters of most similar tuples for data imputation.
Specifically,
in order to impute the missing values of an incomplete tuple $\mathit{t}_{0}$,
CMI uses a kernel-based method to divide the dataset (including both complete and incomplete tuples) into several clusters, 
and use the complete tuples whose plausible values are generated from the same cluster with $\mathit{t}_{0}$ to fill the missing cells.

Fuzzy k-means (FKM) \cite{li2004towards} extends the original k-means clustering method to a fuzzy version for imputing missing data,
which extends the basic idea of partitioning the dataset into several different clusters.
The advantage of applying fuzzy approach is that fuzzy clustering provides a better usage when the clusters are not well-separated, which is a common case in real data imputation application. 
In this process,
the data object cannot be assigned to a concrete cluster represented by a cluster centroid 
(as is done in the basic k-means \cite{macqueen1967}),
because each data object belongs to all clusters with different membership degrees.
FKM attaches each tuple with a membership function which describes the degree that this tuple belongs to the certain cluster. 
Since there are missing data in incomplete tuples, 
we use only complete attribute values to compute the cluster centroids.
The imputation results for the missing values of incomplete tuples are computed based on the information about membership degrees and the values of cluster centroids.

IFC \cite{DBLP:conf/fuzzIEEE/NikfalazarYBK17} designs an iterative fuzzy clustering algorithm to impute missing values,
which further extends the fuzzy clustering method to an iterative version.
After fuzzy clustering, the useful information such as the membership degree and centroids' characteristics are obtained and used to update the previous imputed values.
The the fuzzy clustering will be conducted again over the updated tuples.
This fuzzy clustering and imputation process will perform iteratively,
until the consecutive iterations less than a specified threshold or the number of iterations exceeding the maximum number of iterations.

\subsection{Distance Model based}


Distance model based method LOESS~\cite{Cleveland1996} learns a regression model over similar neighbors, 
which is then used to impute the incomplete data $\mathit{t}_0[\mathit{A}]$ referring to the complete attribute values $\mathit{t}_0[\mathcal{R}\setminus\mathit{U}]$.
Owing to data sparsity and heterogeneity, the neighbors found by complete attribute values $\mathcal{R}\setminus\mathit{U}$ may be distant with each other, and thus do not share the same/similar model. 

Considering the sparsity and heterogeneity problem which usually occurs in the numerical dataset,
IIM \cite{DBLP:conf/icde/ZhangSSW19} proposes to learn an individual model for each complete tuple,
referring to its nearest neighbors.
This idea is similar to the intuition of the conditional dependencies that hold conditionally over a subset of tuples instead of the whole dataset.
For the $k$ neighbors sharing similar values of each tule,
IIM utilizes their individual regression models learned over them to impute missing values.
In order to automatically set the appropriate parameters of various regression models,
IIM adaptively learn the individual model of each complete tuple over a distinct number $\ell$ of learning neighbors.
It determines a proper number $\ell$ and the corresponding model for each tuple,
which can impute most accurately the other complete tuples.
However, IIM can only learn models over the numerical values, 
and cannot handle the categorical data which are often prevalent in the real applications.

Different from IIM,
DLM \cite{kdd20-distanceModel} employs distance likelihood maximization for various incomplete attributes imputation,
which can handle mixed dataset, containing both numerical and categorical data.
Since missing values may appear in various attributes,
including different data types of values and different incomplete attributes in tuples,
there may exist various models having different LHS attributes and RHS attributes in LOESS and IIM with expensive time costs preventing the imputation performing.
DLM proposes to uniformly utilizes the distances on all the attributes with various types of values,
and directly calculate the most likely distances of missing values to avoid enumerating the combinations of imputation candidates and regression models.
By cosidering the consistent distance likelihood computation,
not only the categorical and numerical values can be uniformly processed,
but also the determination and dependent attributes could be imputed simultaneously.

\subsection{Distance Constraint based}
To deal with the similarity relationships between heterogeneous values, 
differential dependencies (DDs) \cite{DBLP:journals/pvldb/SongZC015,DBLP:journals/tkde/SongSZCW20} and matching dependencies (MDs) \cite{DBLP:journals/pvldb/RekatsinasCIR17} are employed for imputation .

In order to tackle the data sparsity and data heterogeneity problems,
where the number of neighbors based on equality relationship is limited,
\cite{DBLP:journals/pvldb/SongZC015} considers similarity neighbors identified by similarity rules with tolerance to small variations to impute missing values.
The enriched similarity neighbors with short distances to incomplete tuples can fill more missing data that are not revealed by the  limited equality neighbors.
Moreover,
\cite{DBLP:journals/pvldb/SongZC015} proposes an approximate imputation approach based on conditional expectation maximization,
which ensures the method achieves a deterministic approximation factor in expectation.

However, \cite{DBLP:journals/pvldb/SongZC015} focuses on imputing single incomplete attribute, 
which is the RHS attribute of the given distance constraint. 
It uses the complete LHS values to find neighbors and infer the missing RHS values referring to the distance rules.
\cite{DBLP:journals/tkde/SongSZCW20} further extends the techniques to general cases for imputing multiple incomplete attributes including LHS attributes of the distance rules. 
It presents the algorithms for handling multiple incomplete attributes, where missing values may occur on the both LHS and RHS attributes of the rules.
Moreover,
\cite{DBLP:journals/tkde/SongSZCW20} also provides an approximation algorithm with the performance guarantee,
which bounds the imputation performance in general cases.


HoloClean \cite{DBLP:journals/pvldb/RekatsinasCIR17} utilizes the distance-based constraint MD to guide the data imputation for incomplete values.
Besides, it also employs the statistical learning and external data as the additional signals to impute missing data.
Given an incomplete dataset,
HoloClean automatically generates a probabilistic program that performs data imputation.
Instead of considering each signal independently,
it unifies various signals for imputation guidance.
To combine different signals, it relies on probability theory to reason about incompleteness across the dataset.
The signals are converted to features of the graphical model and are used to describe the distribution characterizing the given relational instance.
To impute the incomplete values,
HoloClean applies all the available signals including statistical learning and probabilistic inference over the generated model.

\section{Conclusions}
\label{sect:conclusion}

In this article, we provide a comprehensive survey of the state of the art distance-based data cleaning approaches, focusing on four main tasks, including rule profiling, error detection, data repair and data imputation. 
We discuss five frequently-used distance rules, i.e., MDs, MFDs, CDs, FFDs and DDs,
which can not only get a good profiling over the specific dataset, 
but also provide the valuable information to the other three data cleaning tasks. 
Moreover, we summarize the state of the art distance-based error detection approaches into two main categories, 
which can help identify possible erroneous cells for data repair task.
The existing distance-based data repair methods are illustrated with four main classes,
including distance clustering based, distance cost based, distance model based and distance constraint based.
Finally, we review various distance-based data imputation approaches,
which benefit from data repair technology by resolving conflicts and duplications.  

While many traditional data cleaning algorithms have been proposed and implemented in research prototypes and commercial tools, 
further work over distance-based data cleaning is needed, especially in the context of repairing dirty data with the distance signal. 
Moreover, rather than the general data, distance-based cleaning methods dedicated to specific data types are also expected, such as event data \cite{DBLP:conf/icde/ZhuSWYS14,DBLP:journals/tkde/SongGWZWY17}. 
For instance, 
(1) constraints on event distances \cite{DBLP:conf/sigmod/ZhuSL0Z14,DBLP:journals/tkde/GaoSZWLZ18} could be explored to detect errors;
(2) repairing \cite{DBLP:journals/pvldb/0001SZL13,DBLP:journals/tkde/0001SZLS16} and imputation \cite{DBLP:journals/pvldb/SongC016} may thus be performed, e.g., under the constraints on timestamp distances.

\subsection*{Acknowledgement}

This work is supported in part by the National Key Research and Development Plan (2019YFB1705301) and the National Natural Science Foundation of China (61572272, 71690231).

\balance

\bibliographystyle{abbrv}
\bibliography{distance-survey}

\end{document}